\documentclass[reprint,aps]{revtex4-1}

\usepackage{amsmath, amsthm, amssymb, amsfonts}

\usepackage[utf8]{inputenc} 
\usepackage[T1]{fontenc}    
\usepackage{url}            
\usepackage{booktabs}       
\usepackage{nicefrac}       
\usepackage{microtype}      
\usepackage{graphicx}
\usepackage{dcolumn}
\usepackage{bm}
\usepackage{upgreek}
\usepackage{latexsym}
\usepackage{xr}
\usepackage{mathptmx}

\begin{document}

\preprint{APS/123-QED}

\title{Dynamical Transitions of Supercooled Water in Graphene oxide Nanopores: Influence of Surface Hydrophobicity}

\author{Rajasekaran M}
\affiliation{
Department of Chemical Engineering, Indian Institute of Science, Bangalore, India 560012
}
\author{ K. Ganapathy Ayappa}
\email{ayappa@iisc.ac.in}
\altaffiliation{
Centre for Biosystems Science and Engineering, Indian Institute of Science, Bangalore, India 560012
}
\affiliation{
Department of Chemical Engineering, Indian Institute of Science, Bangalore, India 560012
}
  	
\begin{abstract}
Molecular dynamics simulations are carried out to explore the dynamical crossover phenomenon in strongly confined and mildly supercooled water in graphene oxide nanopores. In contrast to studies where confinement is used to study the properties of bulk water, we are interested in the dynamical transitions for strongly confined water in the absence of any bulk-like water. The influence of the physicochemical nature of the graphene oxide surface on the dynamical transitions is investigated by varying the extent of hydrophobicity on the confining surfaces placed at an inter-surface separation of 10 \AA\,. All dynamical quantities show a typical slowing down as the temperature is lowered from 298 to 200 K; however, the nature of the transition is a distinct function of the surface type. Water confined between surfaces consisting of alternating hydrophilic and hydrophobic regions exhibit a strong-to-strong dynamical transition in the diffusion coefficients and rotational relaxation times at a crossover temperature of 237 K and show a fragile-to-strong transition in the $\alpha$-relaxation time at 238 K. The observed crossover temperature is much higher than the freezing point of the SPC/E water model used in this study, indicating that these dynamical transitions can occur with mild supercooling under strong confinement in the absence of bulk-like water. In contrast,  water confined in hydrophilic pore shows a single Arrhenius energy barrier over the entire temperature range. Our results indicate that in addition to confinement, the nature of the surface can play a critical role in determining the dynamical transitions for water upon supercooling. 
\end{abstract}		

\maketitle	
						
\section{Introduction}
The most extensively studied simple liquid, water, continues to fascinate us with its anomalous thermodynamic properties upon supercooling~\cite{speedy1976isothermal,angell1982heat,hare1986densities}. In order to develop a molecular understanding, this anomalous behavior of supercooled water has been extensively investigated by a wide variety of experimental techniques~\cite{clark2010small,dehaoui2015viscosity,smith1999existence}, theoretical approaches~\cite{shi2018origin}, and computer simulations~\cite{saito2018crucial,kumar2007relation}. The homogenous nucleation temperature of water is 235 K at 1 atm~\cite{kanno1975supercooling}, below which liquid water cannot be easily accessed in experiments owing to the formation of ice. However, this so-called "no man's land"  can potentially be accessed by confining water in nanoporous materials or by modifying the freezing point with appropriate additives such as salts, proteins or sugars~\cite{sartor1992calorimetric}. One aspect that has been widely investigated in studies of confined super-cooled water is the presence of the fragile to strong transition (FTS) detected using quasi-elastic neutron scattering, dielectric broadband spectroscopy, and molecular dynamics simulations. Liquids are termed as fragile and strong when the temperature dependence of transport properties follow the Vogel-Fulcher-Tammann (VFT) and Arrhenius temperature dependence respectively. Whether the studies on confined water accurately represent thermodynamic features of bulk supercooled water is an unresolved issue,~\cite{cerveny2016confined} complicated by the degree of confinement, chemical specificity of the confining medium as well as the experimental techniques used to probe the dynamic relaxation upon cooling~\cite{cerveny2016confined}. The degree of confinement, as well as the geometry of confinement, play an essential role in avoiding the formation of a tetrahedral structure, a signature of ice crystallization. Nevertheless, there exists a large body of experimental literature on confined supercooled water in confined media.

Hydrophilic mesoporous silica material MCM-41~\cite{liu2005pressure,wang2015dynamic,wang2015liquid,zhang2011density} has been widely used as a confining matrix in several experimental studies to investigate the properties of supercooled water upon confinement. Water can be supercooled down to $\sim$ 130 K~\cite{liu2013density} upon confinement in MCM-41. Rapid freezing at 232 K is observed in confined water in MCM-41 having cylindrical pores with a diameter of 42 \AA\,. However, in smaller pores of diameter 24 \AA\,, water was found to exist as a  liquid at 160 K in the same material~\cite{morishige1997x}. Quasielastic neutron scattering (QENS) experiments on the supercooled water in MCM-41-S silica pores with a 14 \AA\, diameter~\cite{liu2005pressure} revealed a FTS crossover at 224 K and 1 atm pressure and with the crossover temperature decreasing with increasing pressure up to 1200 bar. Differential scanning calorimetry study on supercooled confined water in calcium silicate hydrates by Zhang et~al.~\cite{zhang2009dynamic} showed a FTS transition at 225 K, concluding that this crossover phenomenon is an inherent nature of water, and that confinement effects do not play a role in this dynamic transition. Further, this crossover phenomenon was attributed to structural changes from a disordered, high-density liquid (HDL) at high temperatures to a low-density, ordered liquid at low temperatures~\cite{wang2015dynamic,mallamace2006fragile}. Using QENS, supercooled water in hydrophobic confinements such as a 14 \AA\, diameter single-wall~\cite{mamontov2006dynamics} and 16 \AA\, diameter double-walled~\cite{chu2007observation} carbon nanotube (CNTs) exhibited a dynamic crossover at 218 and 190 K, respectively. These results suggest that the FTS transition can also be observed in confinement situations where water is mostly in a state of bound water.

In addition to confinement in cylindrically shaped nanopores, studies have also been carried out in layered materials where the confinement geometry allows the distinct layering of water.  Dielectric spectroscopy studies of the relaxation times of water in highly disordered, graphite oxide with interlayer surface spacing ranging  from 6 - 8 \AA\, reveals the presence of an Arrhenius temperature dependence at low temperatures, however only weak signatures of the FTS crossover are observed at 192 K~\cite{cerveny2010dynamics}  which is much lower than the crossover temperatures observed in silica pores~\cite{liu2005pressure,zhang2009dynamic,mallamace2006fragile}. In general, neutron scattering experiments show a stronger and distinct FTS when compared with dielectric spectroscopy data where only a weak transition is observed~\cite{cerveny2010dynamics}.  Water under confinement can co-exist with bulk water, depending on the degree of confinement and levels of hydration, and it is not always easy to separate the contributions arising from bound and free water present in the nanopores. Although molecular dynamics (MD) simulations have been extensively used to study the structure and dynamics of water under confinement~\cite{malani2012relaxation,willcox2017molecular2,kumar2018phase,chakraborty2017confined,kumar2015structure,malani2009influence,cai2019structure}, there have been far fewer studies which focus on supercooled water under confinement. MD simulations in the mildly supercooled regime~\cite{gallo2010dynamic} with confined SPC/E water in silica pores of diameter 15 \AA\, reveal an FTS at 215 K for free water molecules that occupy the interior of the silica pore.  In other MD studies~\cite{kuon2017self} with SPC/E water in cylindrical silica pores having radii ranging between 20 - 40 \AA\, no transitions were observed  in the temperature range 210 K < $T$ < 250 K. 

In contrast to hydrophilic nanopores present in materials such as mica and zeolites, graphene-oxide (GO) based layered materials, consist of nanopores where the extent of hydrophobicity/hydrophilicity can be controlled  during the synthesis of these unique atomically thin materials~\cite{erickson2010determination,xu2017self}. With their nanoporous architecture and increased hydrophilicity when compared with graphite, GO membranes are widely investigated for their potential use in water filtration, gas separations and proton transport applications~\cite{nair2012unimpeded,joshi2014precise,li2013ultrathin,kim2013selective,kim2014high,karim2013graphene}. Several MD studies  have focused on the translational and rotational dynamics of water confined in GO membranes~\cite{willcox2017molecular2,wei2014understanding,raghav2015molecular,devanathan2016molecular,dai2016water} and recently we have investigated the influence of the hydrophobic/hydrophilic ratio as well as the juxtaposition of these different surfaces to form nanopores, on the rotational and translational relaxation of water~\cite{raja2019enhanced}. The inherent mixed hydrophobic/hydrophilic character of the  GO surface results in heterogeneous energy landscapes which can influence the dynamical transitions of  confined water upon supercooling. In this manuscript, we are interested in exploring these dynamical transitions upon supercooling of water confined in GO nanopores with different surface chemistries. 



We carry out molecular dynamics simulations of  SPC/E water confined between graphene oxide (GO) surfaces separated by 10 \AA\, in the temperature range of 200 - 298 K. In this situation, water is present in a layered configuration strongly bound to the surfaces and free or bulk-like water is absent. This interlayer spacing is also found to be optimal for efficient water desalination applications~\cite{joshi2014precise,abraham2017tunable}. The objective of this study is to investigate the dynamical transitions for water confined in a highly inhomogeneous state in the absence of any bulk water. Two types of GO surfaces are investigated to study the influence of the surface chemistry on the dynamics of water. In this study, we consider a GO surface with stripes of oxidized and bare graphene-like regions and  fully oxidized GO surface. The striped GO surfaces are used to create in-registry (IR) nanopore where the oxidized regions on opposing surfaces are in-registry, and fully oxidized (OO) nanopores are formed with the fully oxidized GO surfaces. We analyze in-plane self-diffusion coefficients, self-intermediate scattering functions, and orientational relaxation times. We further check the validity of the Stokes-Einstein relation (SER) in confined water in GO nanopores.  Our results indicate that extent of hydrophilicity on the confining surface plays an important role in determining the nature of the dynamical transitions observed in water confined in GO nanopores.

\section{Molecular dynamics simulation details}

Molecular dynamics (MD) simulations were carried out using LAMMPS~\cite{plimpton1995fast}. Prior to performing confined water simulations, we performed bulk bath simulations in the canonical ($NVT$) ensemble, where the number of particles ($N$), volume of the simulation box ($V$) and temperature ($T$) are fixed, to determine the water loading at various temperatures for the IR and OO nanopores. For bulk bath simulations, the cubic simulation cell of $L$ = 173.5 \AA\, with $\sim$ 175,000 water molecules, was generated with periodic boundary conditions imposed on all three directions and simulation cell consists of a nanopore with the surfaces of $L_{x}$ =  34.12 \AA\, and $L_{y}$ = 29.55 \AA\, separated by a distance $d$ = 10 \AA\,, where $d$ is the distance between opposing carbon atoms on the two surfaces, placed at the center of the simulation cell~\cite{raja2019enhanced}. In order to estimate the size of the simulation cell for bulk bath simulations, we used the local density of water away from the GO surfaces as a measure and ensured that the local density of water along $z$-direction away from the surface approaches the density of bulk water. The trajectory of water molecules was generated from a 15 ns simulation run with a 2 fs time step and was stored at every 0.5 ps time interval. We computed the areal number density, ($\big<N\big>/L_xL_y$) inside the pore from the last 5 ns trajectory and used the number density for the confined water simulations as summarized in Table~\ref{number_density}.

For confined water simulations, the initial configurations with water molecule confined between the surfaces were generated using Packmol~\cite{martinez2009packmol}. Periodic boundary conditions were applied in the lateral directions ($x-y$) with a vacuum above and below the surfaces in the $z$-direction and the periodic box dimensions including the vacuum region are L$_{x}$ = 102 \AA\,, L$_{y}$ = 89 \AA\,, and L$_{z}$ = 100 \AA\,. In our simulations, we treated the surfaces as rigid. Water molecules were modeled using the extended simple point charge (SPC/E) model~\cite{berendsen1987missing} and the bond length and bond angle of a water molecule was constrained using SHAKE algorithm. All-atom optimized potentials for liquid simulation (OPLS-AA) parameters were used for the graphene oxide surfaces along with computed charges~\cite{willcox2017molecular2,wei2014understanding,shih2011understanding,wei2014breakdown,willcox2017molecular}. The van der Waals interactions between GO surfaces and water were modeled using Lennard-Jones 12-6 potential and the cross interaction parameters for LJ potentials were estimated using the Lorentz-Berthelot combining rules. The short-range LJ interactions were smoothly switched to zero between 10 \AA\, and 12 \AA\, and the long-range electrostatic interactions were computed using the particle-particle particle-mesh (PPPM) algorithm~\cite{hockney1980computer}. The equations of motion were integrated using velocity Verlet scheme with a timestep of 2 fs. The system temperature was maintained using the Nos$\acute{e}$-Hoover thermostat with a time constant of 0.1 ps. The system was allowed to equilibrate for 10 - 20 ns followed by a production run of 35 - 200 ns. The trajectories of water molecules were stored at every 0.2 ps to analyze the structural and dynamical properties. Four sets of the trajectories of 40 ps duration at every 4 fs were recorded to obtain the entropy of confined water using the two-phase thermodynamic (2PT) model~\cite{lin2010two}.


\begin{table}[!htpb]
	\caption{Areal number density ($\big<N\big>/L_xL_y$) of water 
	employed in MD simulations for different temperatures, where $\big<N\big>$ is the ensemble averaged number of particles.}
	\centering	
	\begin{ruledtabular}
	\begin{tabular} { c c c }
		T(K) & \multicolumn{2}{c} {Areal number density (\AA$^{-2}$)}\\
		\cline{2-3}
	       	& IR &  OO  \\
					\hline
		200  & 0.1965 & 0.1381  \\
		210  & 0.1936 & 0.1493    \\
		220 & 0.1937  & 0.1611  \\
		230 & 0.1927  & 0.1562   \\
		240 & 0.1911  & 0.1559  \\
		250 & 0.189   & 0.1532 \\
		260 & 0.1868  & 0.152   \\               		
		280 & 0.1847  & 0.151  \\
		298 & 0.1864  & 0.1531  \\
	\end{tabular}
	\end{ruledtabular}
	\label{number_density}	
\end{table}

\begin{figure*}[!htpb]
	\centering
	\includegraphics[scale=0.35]{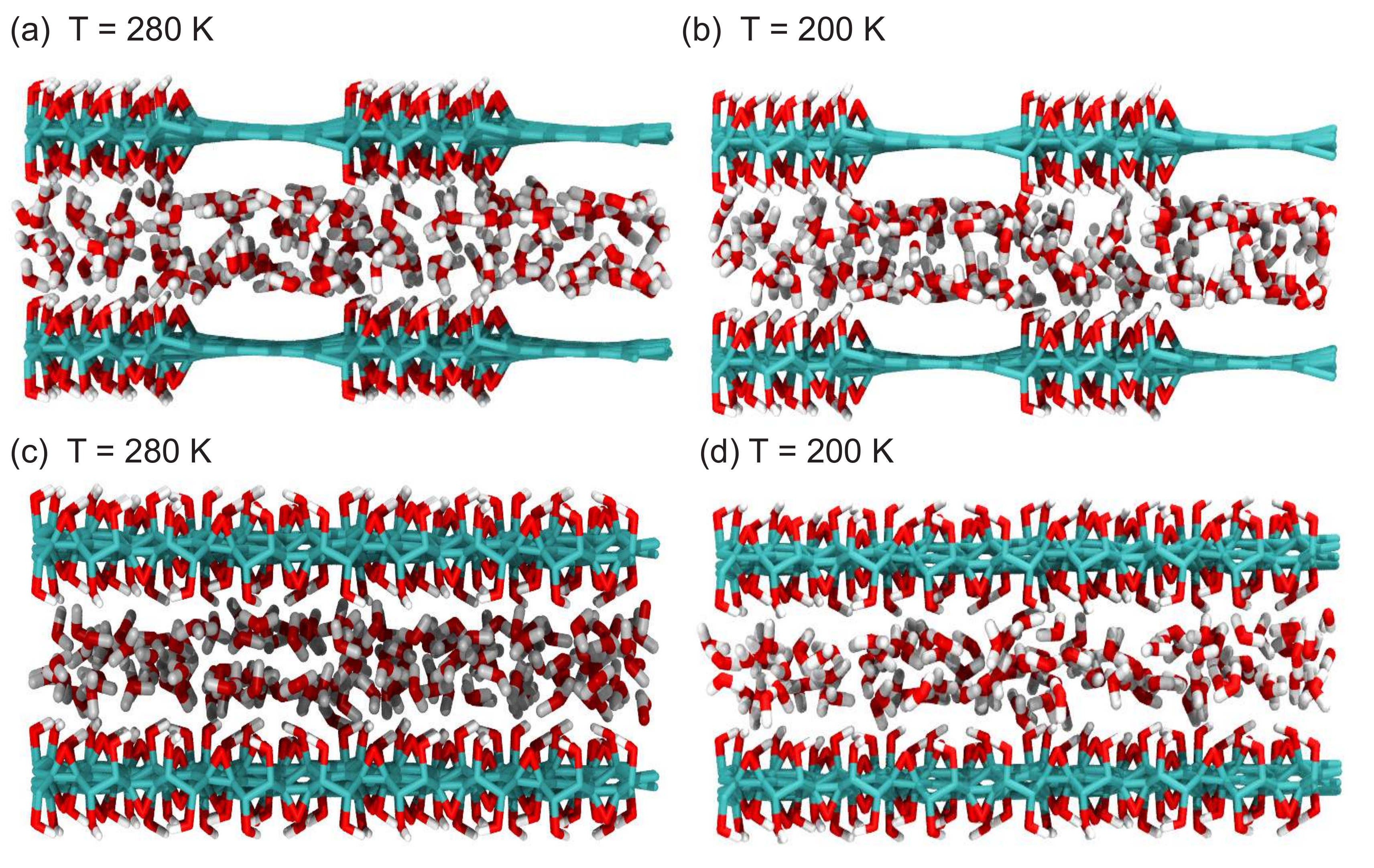}
	\caption {Snapshots depicting water molecules confined between GO surfaces separated at an interlayer separation, $d$ of 10 \AA\, IR pore at (a) $T$ = 280 K, (b) $T$ = 200 K, and OO pore at (c) $T$ = 280 K (d) $T$ = 200 K. Color scheme: carbon - cyan, oxygen - red, hydrogen - white}
	\label{SC_system}
\end{figure*}

\section{Results}

 \subsection{Density distributions}
\begin{figure*}[!htpb]
	\centering
	\includegraphics[scale=0.725]{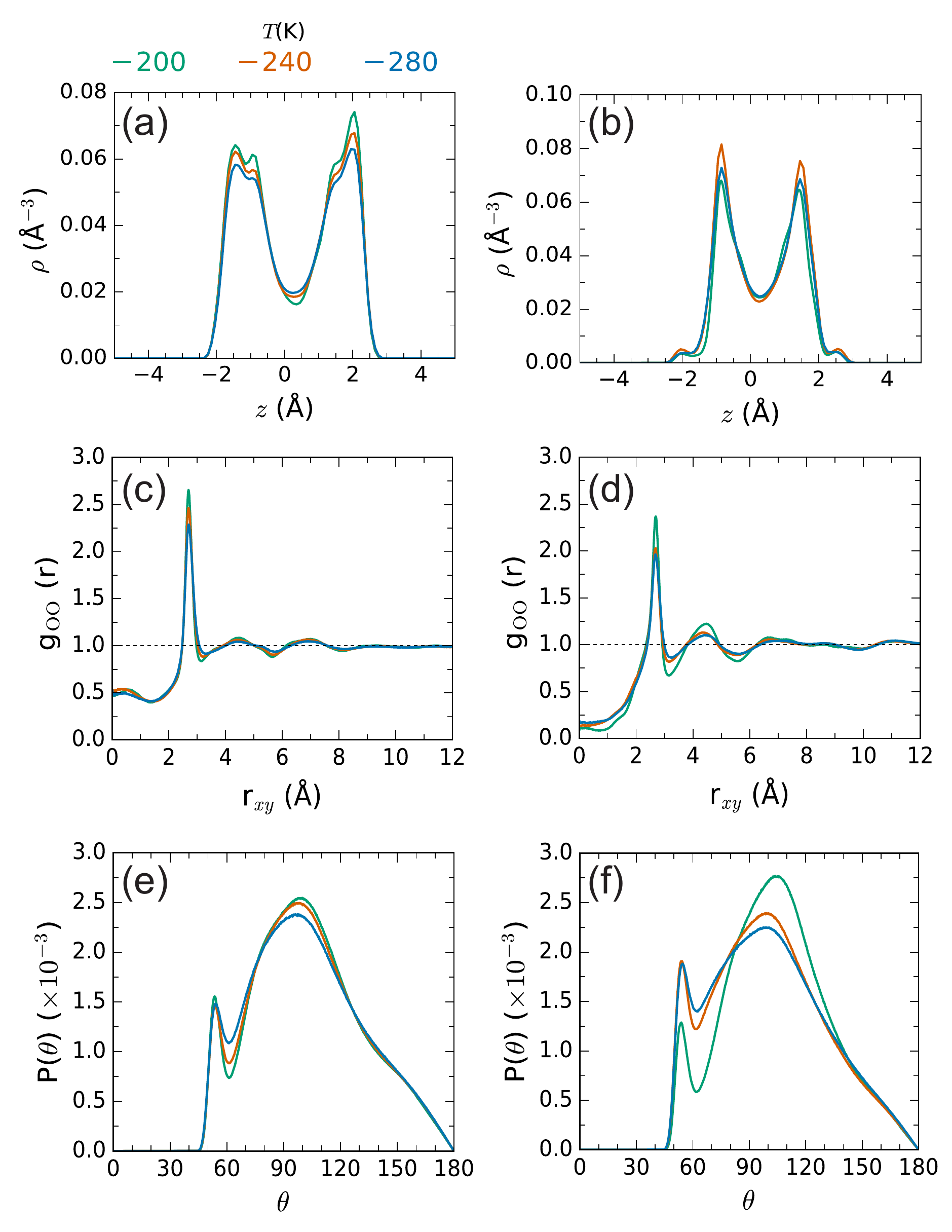}
	\caption {Density profile of water confined in (a) IR pore and (b) OO pore along the $z$ (normal) direction to the GO surfaces. In-plane oxygen-oxygen pair correlation functions of water confined in (c) IR pore (d) OO pore. Distribution of oxygen-oxygen-oxygen angle of water confined in (e) IR pore  and (f) OO pore at various temperatures}
	\label{structure}
\end{figure*}

 We investigate the influence of temperature on layering in GO nanopores by computing the layer-averaged density using,
  \begin{equation} \label{density}\nonumber
 \rho(z) = \frac{\bigg\langle N(z-\frac{\Delta z}{2},z+\frac{\Delta z}{2})\bigg\rangle}{A\Delta z}   ,
 \end{equation}
 where $N(z-\frac{\Delta z}{2},z+\frac{\Delta z}{2})$ is the number of water molecules in a bin of thickness $\Delta z$ in the $z$ direction, $A = L_{x}L_{y}$ , and $\langle .. \rangle $ denotes a time average. The density distribution, $\rho(z)$ illustrated in Figures~\ref{structure}a and b show the presence of two water layers in both the IR and OO pores. The density asymmetry is due to the inherent differences in functionalization present on the GO surfaces~\cite{raja2019enhanced}. For the IR pore, the intensity of the density minima decreases as the temperature is reduced and the interlayer crossing (hydrogen bonding between the two layers) is noticed even at the lowest temperature of 200 K. Small shoulders in the vicinity of density peaks are also observed in both layers due to the striped topology of the IR pore. In case of the OO pore, although two layers are formed the presence of shoulders in the density peaks is absent. The observation of two layers at $d$ = 10 \AA\, in GO nanopores is consistent with earlier studies with the SPC/E water models ~\cite{wei2014understanding,shih2011understanding,willcox2017molecular}. In general, the density peaks are twice that of bulk water (0.033 \AA$^{-3}$), indicating the formation of well-structured water layers in the GO nanopores due to the strong interaction with the hydrophilic GO surfaces.
 
 \subsection{In-plane pair correlation function}
The local structure of water molecules confined between the GO surfaces is investigated by analyzing in-plane pair correlation function (PCF) computed using,
\begin{equation} \label{2drdf}\nonumber
g_{ij}(r) = \Bigg \langle \sum\limits_{i=1}^{N} \sum\limits_{j=1}^{N} \frac{N_{ij}(r+\frac{\Delta r}{2},r - \frac{\Delta r}{2}) A}{2 \pi r \Delta r N_{i} N_{j}} \Bigg \rangle ,
\end{equation}
where $ N_{ij}(r+\frac{\Delta r}{2},r - \frac{\Delta r}{2})$ is the number of atoms $j$ around a atom $i$ at a distance $r$ in a cylindrical shell of thickness $\Delta r$, A = L$_{x}$ L$_{y}$ is the area of simulation box and $N$ is the number of particles in the simulation box. Figures~\ref{structure}c and d illustrate the oxygen-oxygen PCF in GO nanopores for three different temperatures (200, 240 and 280K). The first peak $g_{1}^{max}$ in g$_{O-O}$(r) at $T$ = 200 K is sharper when compared to that at $T$ = 280 K, indicating the increased order at lower temperatures. The increased oscillations in the PCF of the OO pore indicate the increase in ordering of water molecules in the OO pore when compared with the IR pore. The position of first peak $r_{1}^{max}$ in IR and OR pore is 2.69 \AA\, whereas, the peak occurs at 2.75 \AA\,in the SPC/E bulk water, indicating the increased short range order upon confinement. We did not observe any evidence of freezing at the temperatures investigated. The non-zero value of $g(r)$ at $r$ = 0 is owing to projection of molecules in two layers on the $xy$ plane.

\subsection{Distribution of oxygen-oxygen-oxygen triplet angle}
 We analyze the distribution of oxygen-oxygen-oxygen triplet angle within the first hydration shell of water to study local arrangements of water molecules in the IR and OO pores. The oxygen-oxygen-oxygen triplet angle has been shown to be a good measure of the degree of tetrahedral arrangement of water molecules in the confined systems as opposed to the tetrahedral order parameter used for bulk water~\cite{malaspina2010structural,accordino2011comment,biswal2009dynamical}. The O-O-O triplet angle for oxygen atoms are computed only if the O-O distances are within a cutoff distance of 3.35 \AA\, from the oxygen of the reference water molecule. For a perfect tetrahedral arrangement, the angular distribution should exhibit a sharp peak at the angle of 109.5$^\circ$. Figures~\ref{structure}e and f show the angular distribution of oxygen-oxygen-oxygen triplet of water confined in the GO nanopores for three different temperatures (200, 240, and 280 K). We observe two peaks; one at 50$^{\circ}$, and another at 100$^{\circ}$ for all three temperatures in the IR pore (Fig.~\ref{structure}e). The first shorter peak at 50$^{\circ}$ is attributed to interstitial water present in the first hydration shell~\cite{biswal2009dynamical,malaspina2010structural}, and a weak temperature dependence is observed. The broad second peak at around 100$^{\circ}$ signifies a distorted tetrahedral arrangement present in the IR pore, with only a weak intensity increase with decreasing temperature. The water molecules in the OO pore also exhibit two peaks as shown in Fig.~\ref{structure}f. The distribution at 200 K shows a much lower peak intensity at 50$^{\circ}$, and a significantly strong, narrow peak at around 104$^{\circ}$ (Fig.~\ref{structure}f). Similar angular distributions were observed in the hydration water of grooves of DNA~\cite{biswal2009dynamical}.

 \subsection{Translational Dynamics}
\begin{figure*}[!htpb]
 	\centering
 	\includegraphics[scale=0.75]{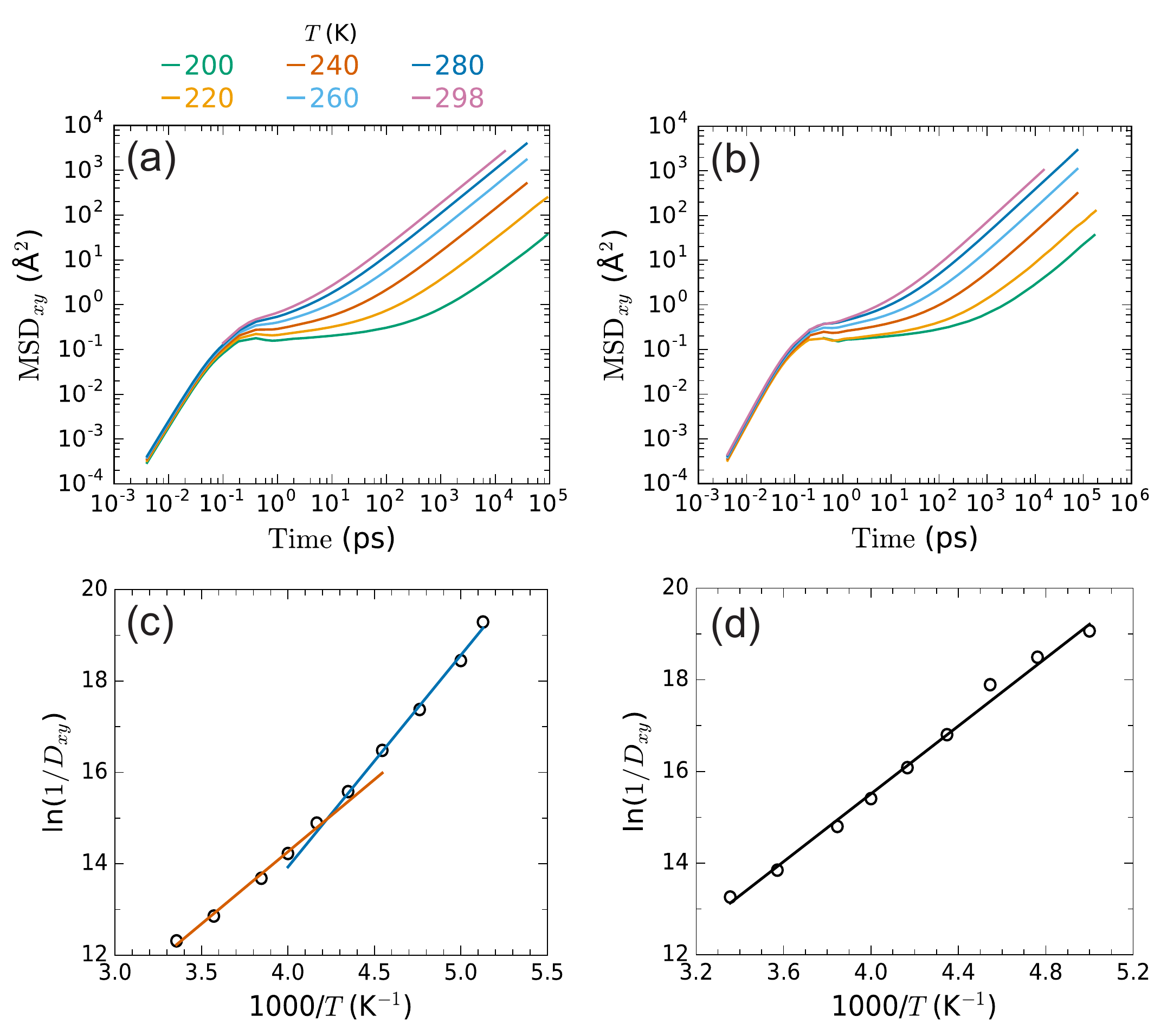}
 	\caption {Time dependence of MSD of water confined in (a) IR pore (b) OO pore at various temperature. Temperature dependence of $D_{xy}$ of water confined in (c) IR pore (d) OO pore. The open circles represent the simulation data, and the solid line shows the fitting data.}
 	\label{msd_1}
 \end{figure*}

 In this section, we investigate the translational dynamics of confined water in the GO nanopores by analyzing in-plane mean squared displacement (MSD$_{xy}$) using,
  \begin{equation}\label{msd_2d}\nonumber
 {\mbox {MSD}}_{xy}(t)  = \Bigg\langle \frac{1}{N} \sum\limits_{i=1}^{N} \vert {x}_{i}(t)-{x}_{i}(0)\vert ^{2} + \vert{y}_{i}(t)-{y}_{i}(0)\vert^{2} \Bigg\rangle_{\tau},
 \end{equation}
where ${x}_{i}(t) $ and ${y}_{i}(t)$ are the $x$ and $y$ coordinates of the center of mass of molecule $i$, $N$ is the total number of molecules, and $\langle ... \rangle _{\tau}$ is the ensemble average over $\tau$ shifted time. Figure~\ref{msd_1}a-b illustrates the time dependence of MSD$_{xy}$ of water confined in IR and OO nanopores at different temperatures. The ballistic regime is observed at short times and the diffusive regime where the MSD$_{xy}$ scales as $t$ is observed at longer times. The onset of caging as evidenced in the plateau regions of the MSD$_{xy}$ at intermediate times occurs at similar times in both the IR and OO nanopores for all temperatures. Extended plateau regimes are observed at lower temperatures similar to the onset of glass-like dynamics. However, the extent of caging is more pronounced in the OO nanopore at lower temperatures (220 - 200 K) due to the restricted hydrophilic environment experienced by water molecules. We note that the MSD$_{xy}$ of confined water in both the IR and OO pores at $T$ = 298 K is much lower than bulk water at $T$ = 298 K (not shown here). The in-plane self-diffusion coefficient ($D_{xy}$) of confined water is computed in the linear regime of the MSD$_{xy}$ using the Einstein relation~\cite{allen2017computer,frenkel2002understanding}. 
  \begin{equation} \label{D2d}
 D_{xy} = \lim_{t\to\infty} \frac{\mbox{MSD}_{xy}(t)}{4t}
 \end{equation}
 The computed $D_{xy}$ values in the IR and OO pores are provided in Table~\ref{table_diffuse}. The translational mobility in the OO pore is significantly lower than that in IR pore for all temperatures. Figure~\ref{msd_1}c-d illustrates the temperature dependence of diffusion coefficient, $D_{xy}$ in the IR and OO pore. In the IR pore (Fig.~\ref{msd_1}c), the inverse of $D_{xy}$ is fitted to the Arrhenius equation, $1/D_{xy} = 1/D_{0}^{A}\exp(E_{A}/RT)$, where $1/D_{0}^{A}$ is the pre-exponential factor, $R$ is the ideal gas constant and $E_{A}$ is the activation energy for diffusion. The fitting of diffusion coefficient data in the temperature range $T$ = 298 - 240 K yields the activation energy, $E_{\text{A}}$ = 26.26 kJ/mol. The diffusion coefficient data in the temperature range, $T$ = 230 - 195 K fits well with the Arrhenius equation with the activation energy, $E_{\text{A}}$ = 38.54 kJ/mol. The higher activation energy at a lower temperatures indicates the formation of a stronger hydrogen bond network in the system. Thus the diffusivity shows a distinct strong-to-strong transition at $T_{\text{C}}$ = 237 K in the IR pore (Fig.~\ref{msd_1}c). In contrast for the OO pore of $d$ = 10\AA\, (Fig.~\ref{msd_1}d) the diffusion coefficient data in the temperature range $T$ = 298 - 200 K show a single Arrhenius behaviour with the activation energy, $E_{\text{A}}$ = 30.75 kJ/mol.
 
\begin{table}[h]
	\caption{In-plane self diffusion coefficients, $D_{xy}$ 
	of confined water in the GO nanopores, estimated from fitting the linear regime of the MSD using the Einstein relation (Eq.~\ref{D2d}).}
	\centering
	\begin{ruledtabular}
	\begin{tabular}{ c c c }
		$T$ (K) & \multicolumn{2}{c} {$D_{xy}$ (cm$^{2}$ s$^{-1}$)}  \\	 
		     \cline{2-3}
		     	& IR  & OO  \\
		\hline
		         195 & 4.185 x 10$^{-9}$  & -  \\
	  	    	 200 & 9.7375 x 10$^{-9}$ & 5.245 x 10$^{-9}$ \\  
	         	 210 & 2.8425 x 10$^{-8}$ &  9.305 x 10$^{-9}$ \\
	        	 220 & 6.955 x 10$^{-8}$ & 1.6955 x 10$^{-8}$ \\
	        	 230 & 1.7145 x 10$^{-7}$ & 5.0425 x 10$^{-8}$ \\	           
	         	 240 & 3.4125 x 10$^{-7}$ & 1.035 x 10$^{-7}$ \\
	        	 250 & 6.645  x 10$^{-7}$ & 2.03475 x 10$^{-7}$ \\
	        	 260 & 1.14275 x 10$^{-6}$ &  3.735 x 10$^{-7}$ \\
	        	 280 & 2.6125 x 10$^{-6}$ & 9.6875 x 10$^{-7}$ \\
	        	 298 & 4.495 x 10$^{-6}$  & 1.74175 x 10$^{-6}$ \\
	\end{tabular}
	\end{ruledtabular}
	\label{table_diffuse}	
\end{table}



 \subsection{Self-intermediate scattering function}
The translational relaxation of water confined in the GO nanopores is studied by analyzing the self-intermediate scattering function, F$_{s}(k,t)$ for the oxygen atom of the water evaluated using,
 \begin{equation} \label{fskt}\nonumber
 \text{F}_{s}(k,t)  = \Bigg\langle \frac{1}{N} \sum\limits_{j=1}^{N} \exp(i\textbf{k}\cdot(\textbf{r}_{j}(t)-\textbf{r}_{j}(0))\Bigg\rangle_{\tau},
 \end{equation}
where $k$ is the wavenumber and $\textbf{r}_{j}(t)$ is the position of the oxygen atom of water molecule. The F$_{s}(k,t)$ is the self part of the Fourier transform of the particle density-density time correlation function~\cite{hansen1990theory} which can be experimentally measured by the quasielastic neutron scattering (QENS) technique~\cite{mamontov2005observation,mamontov2006dynamics,chen2006experimental}. The diffusion coefficient can also be determined from the F$_{s}(k,t)$ evaluated at small wavenumber where the motion of molecules are diffusive~\cite{zanotti1999relaxational}. 

\begin{figure*}[!htpb]
	\centering
	\includegraphics[scale=0.75]{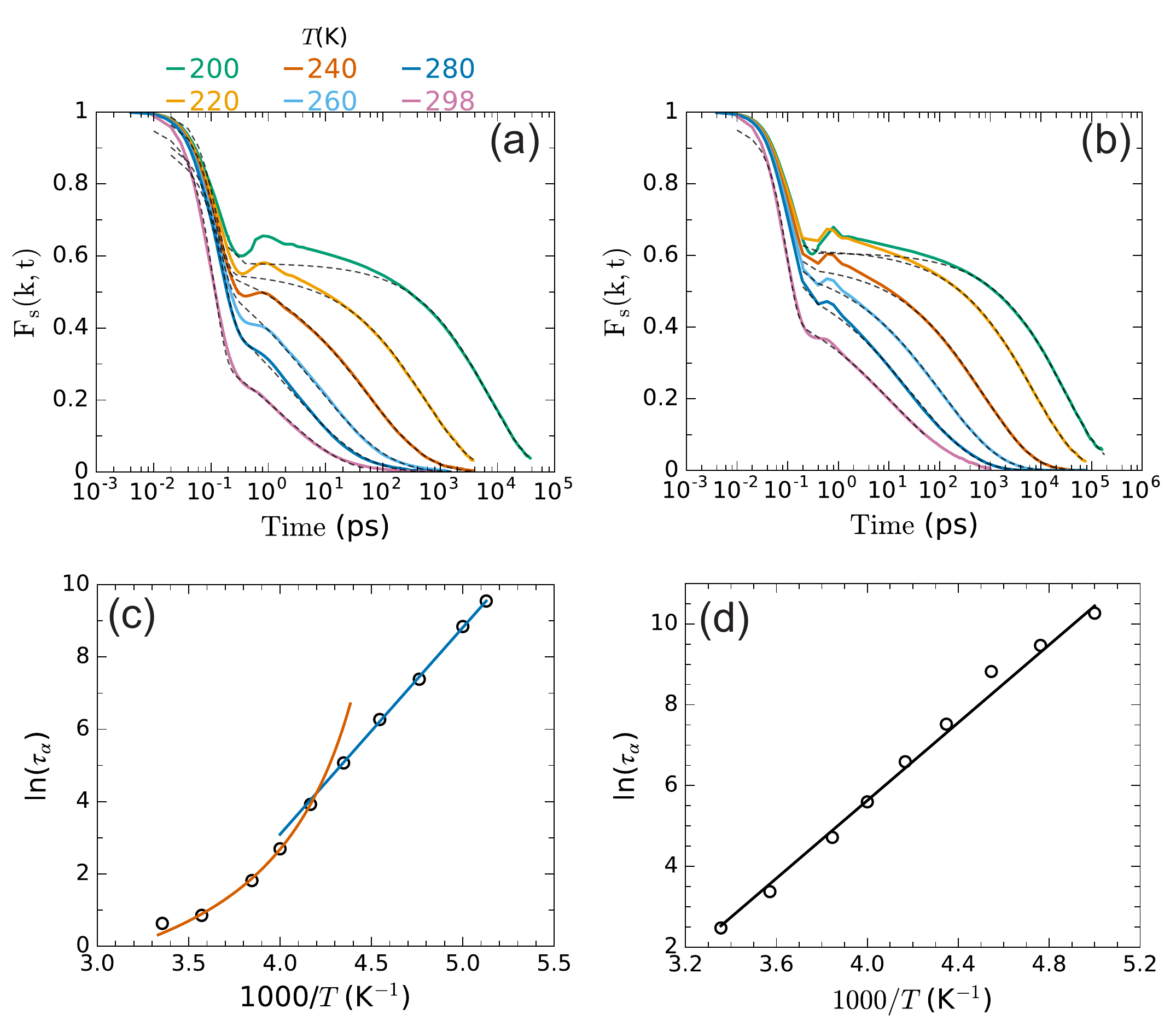}
	\caption { Self-intermediate scattering function, F$_{s}(k,t)$ of water confined in (a) IR pore and (b) OO pore at the wave number, $k$ corresponding to the first peak in the radial distribution function for various temperatures. The solid and dotted lines denote the simulation and model data respectively. The model data is obtained by fitting F$_{s}(k,t)$ to Eq.~\ref{fit_fskt}. Temperature dependence of translational relaxation time, $\uptau_{\alpha}$ of water confined in (c) IR  pore and (d) OO pore. The open circles represent the simulation data, and the solid line shows the fitting data. The dynamic crossover from a fragile to strong liquid is observed in the IR pore.}
	\label{fskt_1}
\end{figure*}

Figure~\ref{fskt_1}a-b shows the time dependence of F$_{s}(k,t)$ of water confined in the IR and OO nanopores of $d$ = 10 \AA\, for temperatures ranging from 298 K to 200 K at the wavenumber that corresponds to the first peak of the radial distribution function. The F$_{s}(k,t)$ exhibits an initial rapid decay, indicating the ballistic regime as noticed in the MSD as well, followed by a weak $\beta$-relaxation at intermediate times, indicating cage motion and then long $\alpha$-relaxation at long times, characterizing the cage relaxation. With decreasing temperature, the plateau which is indicative of the onset of caging becomes more pronounced, and molecules take longer to exit their cages as the temperature is reduced. This caging effect in the OO pore is significantly higher than that in the IR pore. The appearance of a distinct Boson peak which is associated with localized vibrations in the cage is noticed in the IR and OO pores from the temperature of 220  and 240 K. This peak is typically observed in glassy systems~\cite{jakse2008dynamic}, and supercooled water~\cite{gallo2012mode}. In both pores, the $\alpha$-relaxation which is a signature of diffusive motion, slows with decrease in temperature. A stretched exponential function can describe the non-exponential extended $\alpha$-relaxation observed in both the pores for all temperatures. We model the simulated F$_{s}(k,t)$ using~\cite{gallo2012mode,corradini2013microscopic}, 
 \begin{equation}\label{fit_fskt}
 \text{F}_{s}(k,t) = [1-a] \exp\left [-(t/t_{short})^2 \right ]+ a\exp \left[ -(t/\uptau_{\alpha})^\beta \right],  
 \end{equation}
where $a$ is the Debye-Waller factor related to the cage radius, also called as Lamb-M$\ddot{o}$ssbauer factor for the single-particle motion, $t_{short}$ is the short relaxation time, $\uptau_{\alpha}$ is the $\alpha$-relaxation time, and $\beta$ is the stretched exponent, also known as Kohlrausch exponent which ranges from 0 to 1. Gaussian function describes the initial fast decay and the stretched exponential function, known as the Kohlrausch-Williams-Watts (KWW) function represents the slow $\alpha$-relaxation. The model parameters obtained from the fits of F$_{s}(k,t)$ of water confined in the IR and OO nanopores for all temperatures studied are given in Table~\ref{table_fskt}. The fits for the simulated F$_{s}(k,t)$ are obtained using MATLAB 9.1. The $\alpha$ relaxation time, $\uptau_{\alpha}$ of water molecules in the OO pore is always greater when compared with the IR pore for all temperatures, indicating the greater frustration experienced by the molecules on the fully hydrophilic surfaces of the OO pore. In the temperature range of 298 - 210 K, the translational relaxation time of water molecules confined in the OO pore is almost 8 - 20 times higher than that of water molecules in the IR pore, whereas, at $T$ = 200 K, the relaxation time in the OO pore is only twice of that in the IR pore. At much lower temperatures, the relaxation may be less dependent of the pore type. 

Figure~\ref{fskt_1}c-d displays the temperature dependence of the translational relaxation time of water confined in the IR and OO nanopores. In the IR pore (Figure~\ref{fskt_1}c), the translational relaxation time in the temperature range of 298 - 240 K are fitted to a Vogel-Fulcher-Tammann (VFT) equation (fragile behavior),$\uptau_{\alpha} = \uptau_{0}^{\text{VFT}}\exp(BT_{0}/(T-T_{0}))$, where $B$ is the fragility parameter, and $T_{0}$ is the ideal glass-transition temperature with fitting parameters $B$ = 1.01 and $T_{0}$ = 203 K. The low-temperature translational relaxation time data in the temperature range, $T$ = 230 - 195 K are fit with the Arrhenius equation, with the activation, $E_{\text{A}}$ = 47.56 kJ/mol. This activation energy in the IR pore is higher than that of the free water in MCM-41 pore (34 kJ/mol)~\cite{gallo2010dynamic}. The translationl relaxation time of confined water in the IR pore exhibits a dynamic crossover from VFT to Arrhenius behavior at the crossover temperature $T_{\text{C}}$ = 238 K  (Fig.~\ref{fskt_1}c). The $\alpha$ relaxation time data in the temperature range, $T$ = 298 - 200 K for OO pore are fit with the Arrhenius equation with the activation energy, $E_{\text{A}}$  = 40.03 kJ/mol (Fig.~\ref{fskt_1}d). The trend observed for the OO pores is in sharp contrast to the FTS transition observed for the IR surfaces indicating that surface chemistry and texturing play an important role in the supercooled dynamics of strongly confined water.  
 
\begin{table}[!htpb]
	\caption{Model parameters corresponding to the F$_{s}(k,t)$	of water molecules confined in the IR and OO pores of $d$ = 10 \AA\, at various temperatures. The F$_{s}(k,t)$ is fitted to Eq.~\ref{fit_fskt}. The corresponding wave number $k$ is obtained from the radial distribution function.}
	\centering
	\begin{ruledtabular}
	\begin{tabular}{ c c c c c c c }
		Pore type & $T$ (K) & $k$ (\AA$^{-1}$) & a  & $\beta$ & $\uptau_{\alpha}$ (ps) & $t_{short}$ (ps) \\
				\hline
		IR      & 195 & 2.333  & 0.6243  & 0.5514  & 14000 & 0.1196 \\
		        & 200 & 2.333  & 0.5806  & 0.5915  & 6918  & 0.1337 \\
		        & 210 & 2.333  & 0.6428  & 0.4211  & 1612  & 0.1078 \\
	          & 220 & 2.333  & 0.5556  & 0.5177  & 528.7 & 0.1254 \\
	          & 230 & 2.333  & 0.6013  & 0.4219  & 159   & 0.1164 \\	           
				    & 240 & 2.333  & 0.6019  & 0.4031  & 50.6  & 0.1194 \\
			    	& 250 & 2.333  & 0.6687  & 0.3505  & 14.8  & 0.1228 \\
			    	& 260 & 2.333  & 0.6773  & 0.3368 & 6.149 & 0.1396 \\
			    	& 280 & 2.333  & 0.6254  & 0.3351 & 2.349 & 0.1671 \\
			    	& 298 & 2.283  & 0.4204  & 0.4007 & 1.883 & 0.1139 \\

		\\
		OO      & 200 & 2.333 & 0.6084 & 0.5257 & 28710 & 0.01412 \\
		        & 210 & 2.333 &  0.554 & 0.5519 & 12940 & 0.1343 \\
			    	& 220 & 2.333 & 0.6163 & 0.4796 & 6784 &  0.112 \\
			    	& 230 & 2.333 & 0.5858 & 0.458  & 1842 & 0.1165 \\
			    	& 240 & 2.333 & 0.5774 & 0.4398 & 727.2 & 0.1159 \\	
			    	& 250 & 2.333 & 0.5857 & 0.4045 & 269.5 & 0.1109 \\	
            & 260 & 2.333 & 0.5862 & 0.3826 & 111.8 & 0.1146 \\  
            & 280 & 2.333 & 0.5759 & 0.3572 & 29.22 & 0.1201 \\
            & 298 & 2.283 & 0.5117 & 0.3378 & 11.89 & 0.09715 \\

	\end{tabular}
	\end{ruledtabular}
	\label{table_fskt}
\end{table}

 \subsection{Rotational dynamics}
\begin{figure*}[!htpb]
  	\centering
  	\includegraphics[scale=0.75]{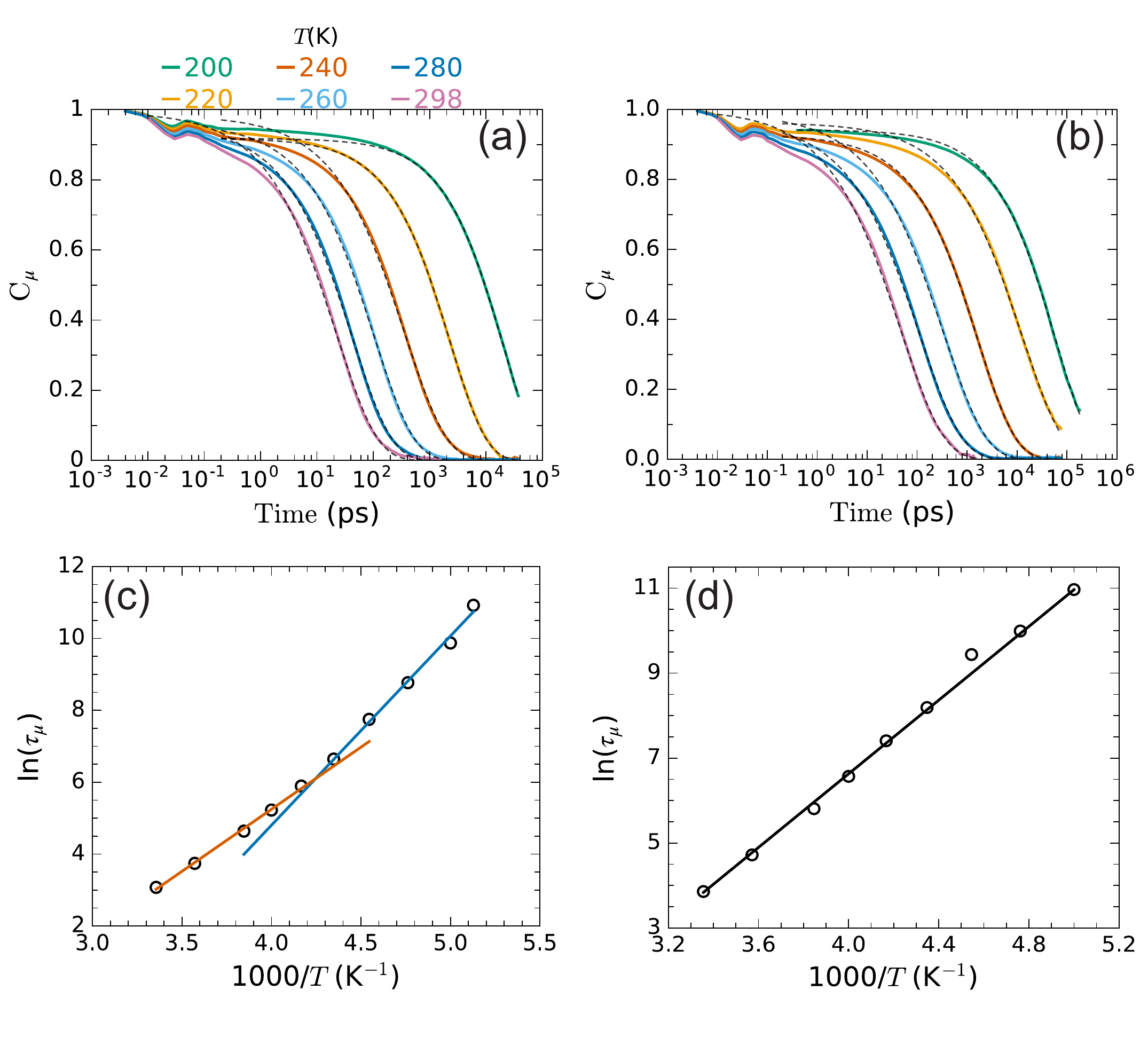}
  	\caption {Dipole-dipole correlation function, $C_{\mu}$ of water confined in (a) IR  and (b) OO pore at different temperatures. The solid and dotted lines represent the simulation and model data respectively. The model data is obtained by fitting $C_{\mu}$ to Eq.~\ref{fit_ocf}. Temperature dependence of dipole relaxation time of water confined in (c) IR pore  and (d) OO pore. The open circles indicate the simulation data, and the solid line shows the fitting data. }
  	\label{dipole_1}
  \end{figure*}

 The rotational dynamics of confined water in the GO pores is investigated by monitoring the dipole autocorelation function computed using,
  \begin{equation} \label{ocf} \nonumber
  C_{\mu}(t) = \Bigg \langle \frac{1}{N}\sum\limits_{i=1}^{N} (\textbf{e}^{\mu}_{i}(t) \cdot \textbf{e}^{\mu}_{i}(0)) \Bigg \rangle_{\tau},
  \end{equation}
where $\textbf{e}^{\mu}_{i}$ is the dipole moment unit vector of the water molecule $i$ in the molecular frame. Figure~\ref{dipole_1}a-b illustrates the time dependence of dipole autocorrelation function of water confined in GO nanopores of $d$ = 10 \AA\, at different temperatures. The dipole autocorrelation function exhibits an initial fast relaxation followed by a slow, long relaxation. Upon lowering the temperature, no change in the initial fast decay is observed, however, a significantly slow decay in the long-time relaxation extending up to the time window of 200 ns is observed in both pores. The $C_{\mu}(t)$is modeled using a stretched exponential function, also known as the Kohlrausch-Williams-Watts (KWW) function ~\cite{biswal2009dynamical,kumar2006molecular},
   \begin{equation}\label{fit_ocf}
    C_{\mu}(t) =  a\exp \left[ -(t/\uptau_{\mu})^\beta \right], 
   \end{equation}    
where the fitting parameters $\beta$ and $\uptau_{\mu}$ are exponent and relaxation time respectively. The value of $\beta$ ranges from 0 to 1. When $\beta$ = 1, stretched exponential becomes a simple exponential form, representing the Debye-like exponential decay. The model parameters for the fit of $C_{\mu}$ is provided in Table~\ref{dipole_1st_SE}. Figure~\ref{dipole_1}c-d displays the temperature dependence of rotational relaxation time of water confined in IR and OO nanopore. In the IR pore (Fig.~\ref{dipole_1}c), the dipole relaxation time in the temperature range of 298 - 240 K is fitted to the Arrhenius equation with the activation energy, $E_{\text{A}}$ = 28.71 kJ/mol. The low-temperature dipole relaxation time data in the temperature range of 230 - 195 K are fit to Arrhenius equation with the activation energy, $E_{\text{A}}$ = 43.76 kJ/mol. The dynamic crossover of rotational motion from Arrhenius to Arrhenius behavior with different slopes (strong-to-strong liquid) is observed at the temperature of $T_{\text{C}}$ = 236 K for confined water in the IR pore (Fig.~\ref{dipole_1}c). The dipole relaxation time data in the temperature range of 200 - 298 K for the OO pore are fitted with an Arrhenius equation with activation energy, $E_{\text{A}}$  = 36.04 kJ/mol (Fig.~\ref{dipole_1}d). Water confined in the OO pore might exhibit a dynamic crossover at lower temperatures for the OO pore. It is noteworthy that the activation energy of rotational motion, $E_{\text{A}}$ is 50.17 kJ/mol from the dipole relaxation experiment on a monolayer of water confined in the GO pore of 8 \AA\,~\cite{cerveny2010dynamics}. This activation energy is much higher than that of the IR and OO pore since a monolayer of water strongly interacts with GO surfaces, and the rotational motion is severely hindered at this lower interlayer spacing~\cite{raja2019enhanced} when compared with the larger spacing of 10 \AA\, considered in this study..

\begin{table}[!htpb]
	\caption{Model parameters corresponding to the $C_{\mu}$ 
	of water molecules confined in the IR and OO pores of $d$ = 10 \AA\, for various temperatures. The $C_{\mu}$ is fitted to Eq.~\ref{fit_ocf}.} 
	\centering
	\begin{ruledtabular}
	\begin{tabular}{  c c c c c  }
		Pore type & $T$ (K)  & a  & $\beta$ & $\uptau_{\mu}$ (ps)  \\	 
			\hline	
		IR  & 195 & 0.9182  & 0.6935   & 55190 \\
		    & 200 &  0.9178  & 0.7077  & 19350  \\
		    & 210 &  0.9363  & 0.6455  &  6410 \\
    		& 220 &  0.9183  & 0.6773  & 2312  \\
    		& 230 &  0.9553  & 0.6354  & 798.3 \\
    		& 240 &  0.9831  & 0.5836  & 361.7  \\
    		& 250 &  0.9377  & 0.6614  & 185.5 \\
    		& 260 &  0.9618  & 0.6116 & 103.3  \\
    		& 280 &  0.9544  & 0.6183 & 42.14  \\
    		& 298 &  0.9932  & 0.5923 & 21.61   \\
		\\
		OO  & 200 &  0.9419 & 0.6072 & 57960  \\
		    & 210 &  0.9113 & 0.602  & 21850 \\
	        & 220  & 0.9623 & 0.527 & 12550   \\
	        & 230  & 0.962  & 0.5392 & 3605  \\
    	 	& 240 &  0.9275 & 0.575 & 1647  \\	
    	 	& 250 &  0.9622 & 0.5392 & 712.8 \\	
    		& 260 &  0.988 & 0.5103 & 333  \\  
    		& 280 &  0.9701 & 0.5151 & 112.2  \\
    		& 298 &  1.0  & 0.5065 & 47.35 \\ 

	\end{tabular}
	\end{ruledtabular}
	\label{dipole_1st_SE}
\end{table}


\subsection{Break-down of Stokes-Einstein relation}
The relationship between viscosity $\eta$ and self-diffusion coefficient $D$ is described by the Stokes-Einstein relation (SER), which is valid for liquids at higher temperatures. However, the SER breaks down at low temperatures, below the critical temperature $T_{C}$ predicted by ideal mode coupling theory. The SER is, 
\begin{equation}\label{SER}\nonumber
\eta =  \frac{k_{B}T}{6\pi r}\frac{1}{D}  
\end{equation}
where $k_{B}$ is the Boltzmann constant, $T$ is the temperature of the liquid, and $r$ is the radius of the particle.
Since the viscosity is proportional to the structural (translational) relaxation time, the SER can be expressed as D $\propto$ ($\uptau_{\alpha}$/$T$)$^{-1}$. In Fig.~\ref{SER}a-b, the self-diffusion coefficient is plotted as a function of structural relaxation time in log-log scale. We investigate the validity of the SER relation for confined water in GO nanopores for the temperature range of $T$ = 200 - 298 K. It is noteworthy that the temperature range considered in this study includes the temperature higher than the melting temperature of SPC/E water model ($T_{m}$ = 215 K~\cite{vega2005melting}). According to the SER relation, the exponent $\xi$ from the fit of D $\propto$ ($\uptau_{\alpha}$/$T$)$^{-\xi}$ should be unity if the fluid follows the SER. The exponent $\xi$ obtained for confined water in the IR pore at $T$ = 298 - 260 K is close to 1 (0.97) (Fig.~\ref{SER}a), indicating the applicability of SER. However, the exponent obtained for $T$ $<$ 260 K is 0.71 (Fig.~\ref{SER}a), indicating a significant deviation from the SER. The exponent for confined water in the OO pore for the entire temperature range studied (200 $\geq$ $T$ $\leq$ 298 K) is 0.73 (Fig.~\ref{SER}b), shows a departure from the SER. For bulk water, the exponent $\xi$ is 0.8 for the ST2 water model~\cite{becker2006fractional}, 0.75 for the mW water model~\cite{limmer2013fluctuations}, 0.84 for the SPC/E water model~\cite{mazza2007connection} and 0.8 for water~\cite{dehaoui2015viscosity}. The deeply supercooled bulk TIP4P/2005 water exhibits the breakdown of Stokes-Einstein relation with exponent of 0.74~\cite{saito2018crucial}. As shown in Fig.~\ref{SER}c, the D$\uptau_{\alpha}$/$T$ is constant in a temperature range $T$ = 260- 298 K, and increases drastically with decreasing temperature in the IR pore. For the OO pore, the D$\uptau_{\alpha}$/$T$ increases with decrease in temperature in the temperature range (298 - 200 K). A crossover from the SER ($\xi$ = 1) to fractional SER ($\xi$ = 0.71) for confined water in the IR pore occurs at $T$ = 260 K (Fig.~\ref{SER}c), which is $\sim$ 1.2$T_{m}$ in the SPC/E bulk water, and $\sim$ 2$T_{g}$, where $T_{g}$ is the glass transition temperature of bulk water (136 K). This observed breakdown of the Stokes-Einstein relation in confined water can be attributed to the emergence of spatially heterogeneous dynamics which leads to the decoupling of translational diffusion from translational relaxational dynamics~\cite{dehaoui2015viscosity,jaiswal2015atomic,Wei2018PCM}.

\begin{figure*}
	\centering
	\includegraphics[scale=0.75]{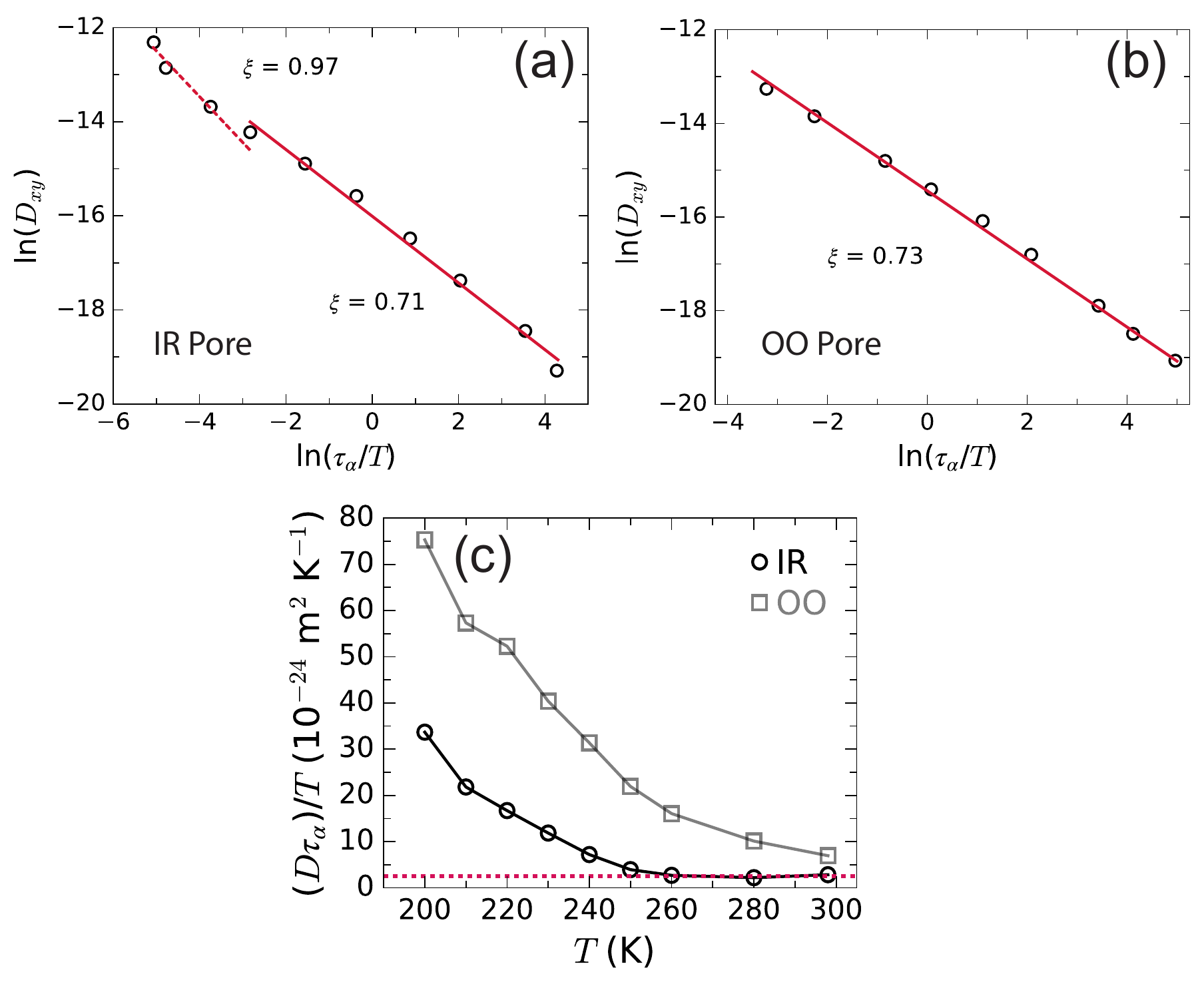}
	\caption {Breakdown of Stokes-Einstein relation in confined water in (a) IR pore (b) OO pore. The diffusion coefficient is fitted to a power law as a function of $\uptau_{\alpha}$/$T$, D$_{xy} \propto$ ($\uptau_{T}$/T)$^{-\xi}$. The exponent, $\xi$ = 0.71 and 0.73 in IR and OO pores shows a significant deviation from Stokes-Einstein relation. (c) Cross-over temperature from Stokes-Einstein to fractional Stokes-Einstein relation. Dotted line denotes a line where D$\uptau_{\alpha}$/$T$ = constant}
	\label{SER}
\end{figure*}

  \begin{figure*}[!htpb]
	\centering
	\includegraphics[scale=0.425]{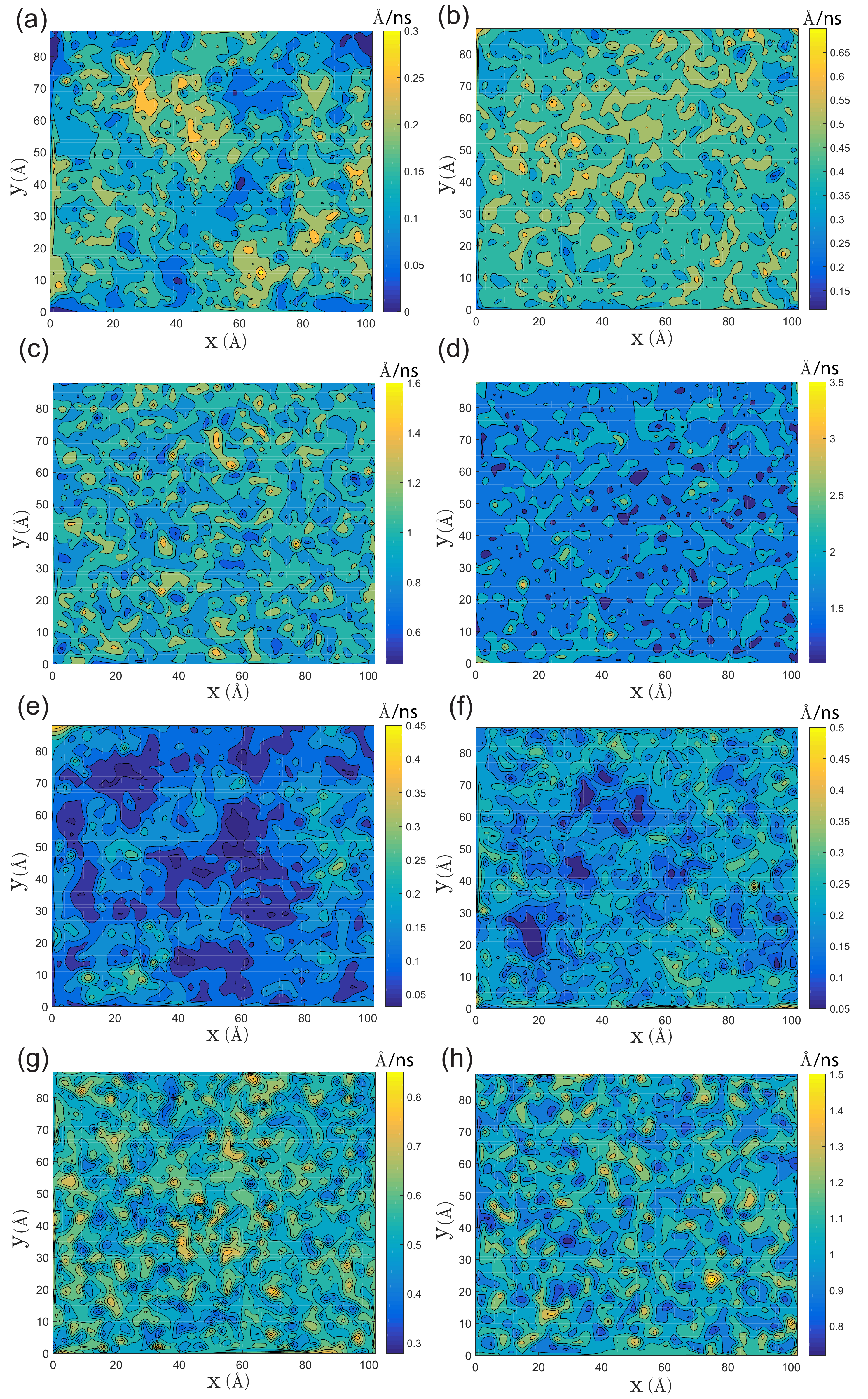}
	\caption {Mobility map of water molecules confined in IR pore at the temperature of (a) $T$ = 200 K, (b) $T$ = 220 K, (c) $T$ = 240 K, and (d) $T$ = 260 K. Mobility map of water molecules confined in OO pore at the temperature of (e) $T$ = 200 K, (f) $T$ = 220 K, (g) $T$ = 240 K, and (h) $T$ = 260 K. These maps are evaluated at the time interval of 10 ns.}
	\label{mobility_map}
\end{figure*}

  \begin{figure*} [!htpb]
	\centering
	\includegraphics[scale=0.625]{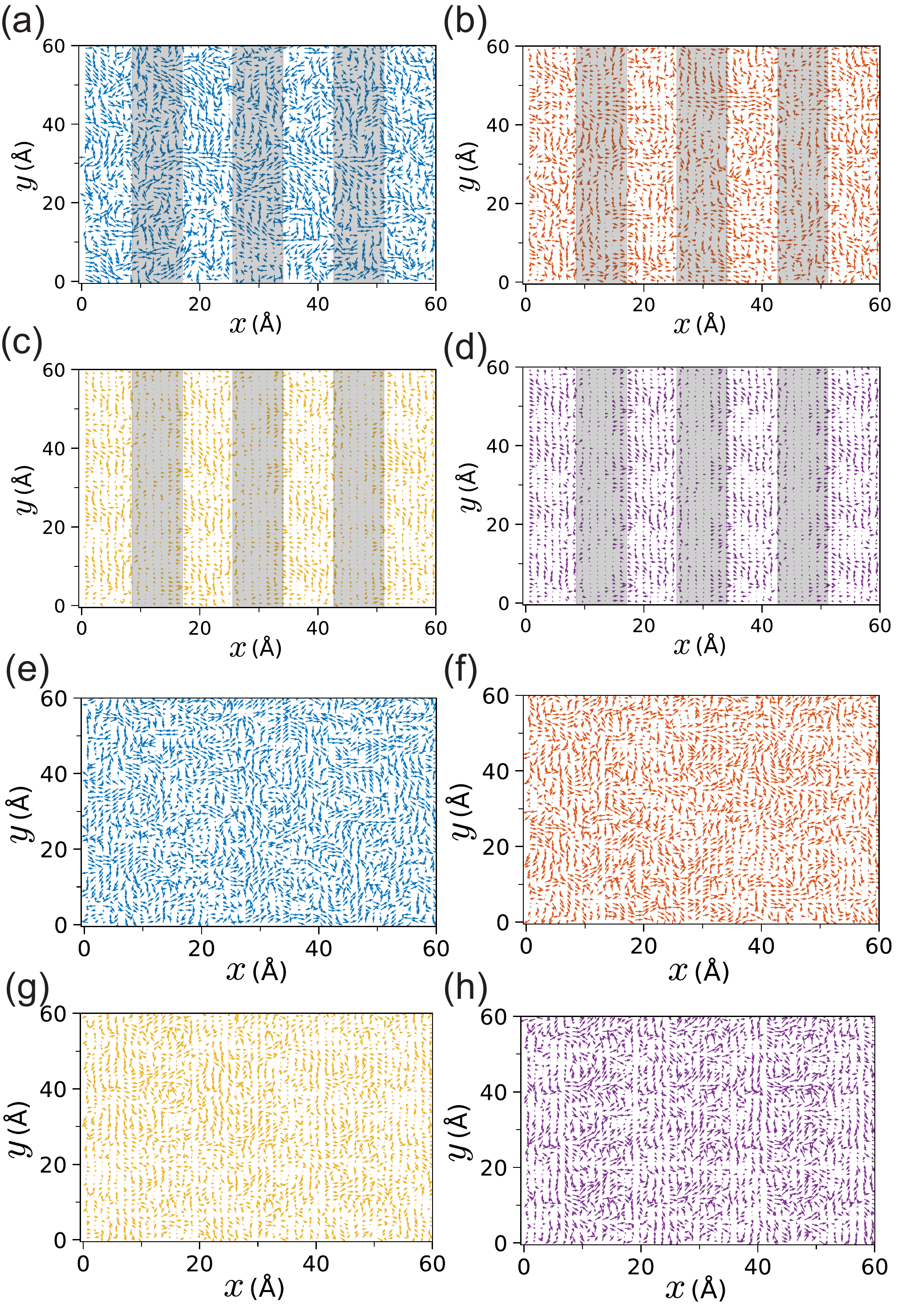}
	\caption {2D dipole map ($x$ - $y$) of water molecules confined in IR pore at the temperature of (a) $T$ = 200 K, (b) $T$ = 220 K, (c) $T$ = 240 K, and (d) $T$ = 260 K. 2D dipole map ($x$ - $y$) of water molecules confined in OO pore at the temperature of (e) $T$ = 200 K, (f) $T$ = 220 K, (g) $T$ = 240 K, and (h) $T$ = 260 K. The water molecules in the layer $z$ > 0 are included for the dipole map computation. The gray color stripes represent the bare graphene-like regions in the IR pore.}
	\label{dipole_map_10}
\end{figure*}

\section{Discussion}
Molecular dynamics simulations have been carried out for water confined between graphene oxide surfaces having different degrees of hydrophobicity. For the IR pores, each surface consists of equally spaced oxidized and bare graphene regions arranged in stripes with both surfaces placed in-registry to form the confining surfaces and in the case of the OO pores, both surfaces are fully oxidized giving rise to a purely hydrophilic confinement. The SPC/E model for water which has been extensively used to study dynamics of confined and supercooled water has been used in this study. The MSDs, F$_{s}(k,t)$ and $C_{\mu}(t)$ all show classic signatures of dynamic slowdown as the temperature is reduced, however differences emerge between the IR and OO pores. Different regimes observed by the dynamical quantities examined are given in Table~\ref{data_summary}.

\begin{table*}[!htpb]	
	\caption{Summary of dynamic crossover of different processes in the IR and OO pores of $d$ = 10 \AA\,}
	\centering
	\begin{ruledtabular}
	\begin{tabular} { c c c c c c }		
Pore & Quantity & $T$ (K) & Fitting & $E_{\text{A}}$ (kJ/mol) & $T_{\text{C}}$ (K) \\ 
	\hline
		 & D$_{xy}$ & 298 - 240 & Arrhenius & 26.26 &  \\
         & D$_{xy}$ & 230 - 195 & Arrhenius & 38.54 & 237 \\
		 & $\uptau_{\alpha}$ & 298 - 240 & VFT & - &  \\
IR       & $\uptau_{\alpha}$ & 230 - 195 & Arrhenius & 47.56 & 238 \\
		 & $\uptau_{\mu}$ & 298 - 240 & Arrhenius & 28.71 &  \\
         & $\uptau_{\mu}$ & 230 - 195 & Arrhenius & 43.76 & 236 \\
\\
		 & D$_{xy}$ & 298 - 200 & Arrhenius & 30.75 & - \\
	OO	 & $\uptau_{\alpha}$ & 298 - 200 & Arrhenius & 40.03 & - \\
	     & $\uptau_{\mu}$ & 298 - 200 & Arrhenius & 36.04 & - \\ 
	\end{tabular}
	\end{ruledtabular}
	\label{data_summary}
\end{table*}

For the IR pores, all dynamical quantities show a distinct crossover temperature at~237 K; however, a single Arrhenius temperature dependence is observed for the OO pores. Interestingly, the type of crossover observed depends on the dynamical quantity being studied. The self-diffusivity dependence on temperature shows a transition from a strong-to-strong transition where $E_{\text{A}}$ in the high temperature regime (240 - 298 K) is 26.26 kJ/mol increasing to 38.54 kJ/mol at lower temperatures with $T_{\text{C}}$ = 237 K (Table~\ref{data_summary}). In  contrast, the relaxation time $\tau_{\alpha}$ obtained from the $\alpha$ relaxation regime of the F$_{s}(k,t)$ shows a distinct FTS transition at $T_{\text{C}}$ = 238 K. The relaxation time $\uptau_{\mu}$ obtained from the dipole-dipole relaxation shows a strong-to-strong transition at $T_{\text{C}}$ = 236 K.

The FTS transition observed for the $\alpha$ relaxation is qualitatively similar to the FTS transition observed by several experimental studies on confined water using QENS experiments. The confinement between graphene oxide surfaces in this study is distinctly different from the confinement situations in MCM and silica pores and more closely resembles the layered confining geometry in clays and graphite oxide materials where the water density is inhomogeneous and strongly influenced by the surface. We first contrast our results with other MD studies where the SPC/E water model has been used to study supercooled water under confinement in large pores where bulk-like water is observed. 

Studies by Gallo and co-workers~\cite{gallo2010dynamic} in the mildly supercooled regime have shown the presence of the FTS transition in the $\uptau$ extracted from the F$_{s}(k,t)$ for bulk-like water confined in cylindrical silica pores. The transition was found to occur at 215 K which occurs at the freezing point of bulk SPC/E water; however, the temperature is very similar to the crossover temperature observed in QENS experiments in silica pores. In our study, with a purely hydrophilic confinement for the OO surfaces, only a single Arrhenius exponent with $E_{\text{A}}$ = 40 kJ/mol describes the dynamics across the entire temperature range for strongly confined water. This activation energy is about 6 kJ/mol higher than the activation energy observed for bulk-like water in the MCM pore. In contrast, the FTS transition occurs at an elevated temperature of 238 K, only in the IR surfaces, a consequence of the increased mobility due to the underlying hydrophilic-hydrophobic environment encountered by the water molecule. Thus our study shows that the fragile behavior of water as reflected in the relaxation time at higher temperatures is driven by the texturing of the confining surface. In other studies using SPC/E water by Ladanyi and co-workers in MCM 41 pores~\cite{kuon2017self}, a monotonic increase in relaxation time was observed with a decrease in temperature. Due to the limited temperature range (210 - 250 K), the FTS transition was not explored. QENS experiments have been extensively used to study the FTS transition in supercooled confined water, and the transition has been observed in a wide variety of materials such as cement paste~\cite{zhang2009dynamic}, carbon nanotubes~\cite{mamontov2006dynamics} and  MCM-41 pores~\cite{wang2015dynamic} where the transition occurs between 218-225 K depending on the material~\cite{swenson2014dynamics}. 

Molecular dynamics simulations of bulk supercooled TIP4P-Ew water~\cite{zhang2009dynamic} exhibits an FTS  transition at  215 K and Biswal et~al.~\cite{biswal2009dynamical} showed that the supercooled hydration water (TIP5P water model) in the grooves of DNA displays an FTS transition (for major grooves) and strong-to-strong (for minor grooves) at 255 K. The strong liquid behavior in the minor grooves was attributed to the increased structuring when compared with water present in the major grooves. This is consistent with the increased structuring of water present between the OO surfaces when compared with the IR surfaces (Figures~\ref{structure}c-f). Thus although an FTS transition observed in water appears to be widely prevalent across both soft and strong confinements, our results reveal that the transition is not observed for highly confined water between purely hydrophilic surfaces and is observed only in the IR pores induced by a mixed hydrophilic-hydrophobic interface which effectively lowers the structuring of confined water. 

Our results with the rotational relaxation dynamics can be connected with results from dielectric spectroscopy used to probe the rotational relaxation modes of confined water. Dielectric relaxation data for water confined in graphite oxide~\cite{cerveny2010dynamics} with hydration dependent interlayer spacing (5.7 \AA\, for dry graphite and 7.9 \AA\, for 25 wt\% water) reveal an Arrhenius behavior at low temperatures; however, the FTS transition observed in QENS experiments is not observed. Dielectric relaxation data on Na-vermiculite clays~\cite{bergman2000dielectric} where intercalated water is found to form two layers of 6 \AA\, thickness between the platelets were found to show a single Arrhenius temperature dependence in the deeply supercooled region (125 - 215 K). Although we have not explored the deeply supercooled regime for the SPC/E water model, the temperature dependence for the dipole-dipole relaxation of confined water with the OO pores, which sample very similar levels of confinement as the clays samples, are consistent with the Arrhenius temperature dependence. Although water is not frozen at the lowest temperature of 195 K, water is highly structured, and a long plateau regime is observed in the MSD$_{xy}$ indicative of a caging regime extending over three orders in time. Similar strong liquid behavior has been observed by Biswal et~al.~\cite{biswal2009dynamical} who show that the rotational dynamics of supercooled hydration water (TIP5P water model) in the major and minor grooves of DNA display a strong-to-strong liquid transition. 

Our results indicate that the presence of an FTS transition occurs only for $\alpha$ relaxation for the IR surfaces, and a strong-to-strong transition is observed for both the diffusion coefficients and the rotational relaxation times. Extended simulations to higher temperatures of 340 K provided additional confirmation for the strong behavior. Interestingly, the crossover temperature is similar for all three dynamical quantities (Table~\ref{data_summary}). At high temperatures, the displacement of a particle is influenced by the alternating hydrophilic and hydrophobic stripes in the IR pores. The presence of the graphene surface significantly enhances the mobility of the water molecules when compared with the OO pores. As the temperature is reduced, there is greater extent of slowing down experienced by the water in the IR surfaces when compared with the OO surfaces where water mobility is retarded to begin with at the higher temperatures. These effects are observed in the mobility maps (Fig.~\ref{mobility_map}) and MSD vs. time data (Fig.~\ref{msd_1}a-b) for different surfaces. 
The mobility maps evaluated over a time interval of 10 ns show about a one order decrease in mobility from the high temperature to the low temperature across the transition temperature of 237 K. In contrast, the mobilities for the OO surfaces decrease only by a factor of three across the transition. In the OO surfaces, there is lowered dynamic heterogeneity as revealed in the mobility differences at a given temperature. At higher temperatures (260 K), the domains of different mobility for low, intermediate and high regions (as discerned from the color maps) are evenly populated for both the OO and IR pores. However, at 240 K, there is a distinct emergence of equally populated larger domains in the IR surfaces with regions of intermediate mobility forming a continuous percolating background interspersed with islands of lower mobility. Thus for a particle to relax at the lower temperatures, the energetics associated with the movement of particles is limited by the slower domain,~\cite{krishnan2012glassy} resulting in a higher barrier for relaxation when compared with the barrier at higher temperatures. Below the transition temperature, the low and intermediate mobility regions extend across the sample interspersed with higher mobility disconnected domains. Hence a distinct increase in dynamic heterogeneity is observed across the transition. In the case of the OO pores, the length scale as seen from the size of the domains in the mobility maps is much smaller; however, the low mobility regions appear to continuously expand across the sample, as the temperature is lowered. This lowered dynamic heterogeneity results in an intermediate value of $E_{\text{A}}$ (Table~\ref{data_summary}). These discerning features in the mobility maps yield information on the displacements which connect with the diffusivities on the one hand and if one views the F$_{s}(k,t)$ as density-density correlations in Fourier space the $\alpha$ relaxation is connected with the relaxation of particle displacements for $k$ vectors corresponding to the first neighbor shells. Our results suggest that the particle displacements, as revealed in the diffusivity as a function of temperature, show a strong-to-strong transition. However, the dynamics of the $\alpha$ relaxation are influenced by the domain topology and show a distinct FTS transition, albeit at the same crossover temperature (Table~\ref{data_summary}). We point out however that the diffusivity transitions observed in MD simulations of bulk TIP4P/Ew water~\cite{zhang2009dynamic} show a FTS transition similar to the $\alpha$ relaxation.

Additional insight into the rotational relaxation can be obtained by examining the dipole vector projection ($x-y$) maps (Fig.~\ref{dipole_map_10}) where distinct differences emerge between the two surfaces. These dipole vector maps are obtained from averaging over 25000 molecular configurations. The templating effect in the IR surfaces is clearly observed with water dipoles above the graphene regions adopting a surface normal orientation when compared with water dipoles in the vicinity of the oxidized regions adopting a more surface parallel orientation~\cite{raja2019enhanced} (Figure~\ref{SC_system}). As the temperature is lowered, distinct domains of oriented dipoles emerge, however, increased heterogeneity is observed in the IR surfaces when compared with the OO surfaces. There is a greater mix of oriented domains, as seen by groups of water molecules having similar orientations located within domains of different orientations in the IR surfaces when compared with a reduced amount of orientational heterogeneity in the OO surfaces. Thus the templating effect in the IR surfaces appears to result in a greater degree of frustration for rotational degrees of freedom giving rise to a change in the rotational energy landscape with temperature leading to the observed strong-to-strong transition. This transition is in contrast with the single strong dependence observed for the OO surfaces. 

The Stokes-Einstein relation (SER) offers additional insight into the influence of the different surfaces on the relaxation dynamics. Interestingly for the IR surfaces, SER is valid at higher temperatures (T > 260 K); however at lower temperatures, deviations from the SER are observed, indicating that the increase in viscous relaxation time  
 ($\uptau_{\alpha}$) is larger than the decrease in the diffusion coefficient as the temperature is reduced. In the case of the OO surfaces, the SER breaks down over the entire temperature range studied.

\section{Summary and conclusion}

In conclusion, our study reveals several interesting features on the water relaxation dynamics upon strong confinement. Water confined between the graphene oxide surfaces provide an "effective" interlayer spacing of ~ 6 - 7 \AA\, where water can form two contact layers with interlayer hydrogen bonding.  The physicochemical nature of the surface is found to play an important role in the temperature dependence of water relaxation, resulting in activation energy barriers that depend on the surface. For a purely hydrophilic surface, only a single activation energy barrier is observed, resulting in a strong relaxation behavior over the entire temperature range for all the dynamical quantities studied. However introducing a mixture of hydrophilic and hydrophobic surfaces brings about distinct changes in the crossover behavior.  The diffusion coefficients and rotational relaxation times show a strong-to-strong crossover, and the $\alpha$ relaxation times show a fragile-to-strong crossover. These dynamical transitions are different from the transitions observed for bulk-like water under confinement, indicating that surface effects can significantly modulate water dynamics upon strong confinement. Examining the 2D mobility and dipole orientational maps provide some physical insight into the influence of surface texturing on the water dynamics. Hence water under strong confinement depicts unique signatures of dynamic slowdown upon mild supercooling distinct from the dynamics associated with bulk supercooled water.

\begin{acknowledgements}
We would like to thank the Supercomputer Education and Research Centre (SERC) and the Thematic Unit of Excellence (TUE) in the Solid State and Structural Chemistry Unit (SSCU) for computing resources and the Department of Science and Technology, India for funding. We also thank Chandan Dasgupta for useful discussions on the results reported in this paper.
\end{acknowledgements}

\bibliographystyle{apsrev4-2}
\bibliography{Reference_scw} 

\end{document}